\documentclass{moriond}

\def\Journal#1#2#3#4{{#1} {\bf #2}, #3 (#4)}

\def\PRL{\em Phys. Rev. Lett.}
\def\PRD{{\em Phys. Rev.} D}
\def\PRA{{\em Phys. Rev.} A}

\def\be{\begin{equation}}
\def\ee{\end{equation}}
\def\bea{\begin{eqnarray}}
\def\eea{\end{eqnarray}}

\newcommand{\Photo}{}

\begin{document}
\vspace*{4cm}
\title{MITIGATION OF PARAMETRIC INSTABILITY}

\author{ M. TURCONI, T. HARDER, R. SOULARD, W. CHAIBI }

\address{ARTEMIS, Universit\'e C\^ote d'Azur, CNRS and Observatoire de la C\^ote d'Azur,
Boulevard de l'Observatoire F-06304 Nice, France}

\maketitle\abstracts{
A key action for enhancing the sensitivity of gravitational wave (GW) detectors based on laser interferometry is to increase the laser power. However, in such a high-power regime, a nonlinear optomechanical phenomenon called parametric instability (PI) leads to the amplification of the mirrors vibrational modes preventing the detector functioning. Thus this phenomenon limits the detectors maximum power and so its performances. Our group has started an experimental research program aiming at realizing a flexible and active mitigation system, based on the radiation pressure applied by an auxiliary laser. A summary on the PI mitigation techniques will be presented, we will explain the working principle of the system that we are implementing and report about the first experimental results.}

\section{Introduction}

\subsection{Parametric Instability in Gravitational Wave Detectors}

Gravitational wave detectors based on laser interferometry have reached a sensitivity h = $\Delta$L/L $\approx$ 10$^{-23}$/$\sqrt{\rm{Hz}}$ at about 100Hz which makes detection of sources in 10Hz-10kHz bandwidth such as black holes and neutron stars binaries, possible, opening a new era for astronomy \cite{Abbott2016,Abbott2017b,Abbott2017a,Abbott2018}. Advanced detectors aLIGO and AdVirgo still haven’t reached their nominal performances and improvements of sensitivity are foreseen in order to push the limits of the observable universe and to increase the detection rate. Moreover, since about ten years, researchers work at the design of a third generation GW ground-based detector: the Cosmic Explorer \cite{Abbott2017c} and the Einstein Telescope (ET) \cite{Hild2010}. The ET, a proposed European project, aims at achieving a sensitivity ten times better than the advanced detectors on a broader spectral bandwidth (1-10kHz). A key action for improving the signal to noise ratio is to increase the laser power. The power inside the Fabry-P\'erot cavities composing the interferometer arms of the advance detectors is foreseen to approach 1 MW and 3 MW for ET. This high circulating power is needed to decrease the shot noise level which limits the detector sensitivity at high frequency (above 200 Hz). But increasing the optical power also means to deal with parametric instability.
PI consists in the amplification of the mirror mechanical modes, initially thermally excited, by radiation pressure. This pressure is exerted by the intracavity laser field composed by the main optical mode and one or several high order modes. The latter are created by scattering of the main mode by the mirror vibrations themselves. PI is analogous to Stimulated Brillouin Scattering (SBS\cite{Bai,Valley}), the parametric interaction involved is the conversion of the main laser mode $\omega_0$ into one phonon (acoustic excitation) $\omega_m$ and one photon at lower energy $\omega_s$ = $\omega_0 – \omega_m$. Two conditions need to be met for the instability to occur: all the three modes involved must be resonant in the cavity (for the optical modes) and in the mirror (for the mechanical mode) and there must be a significant spatial overlap between the beat note and the mechanical mode profiles.\\
The first one-dimensional analysis of PI in the context of advanced GW laser interferometers detectors was described by Braginsky \cite{Braginsky2001}. Since then, several theoretical and experimental studies have been carried-out in order to understand the PI impact on the gravitational wave detectors $^{16-24}$. This impact was eventually observed in 2015 at aLIGO and reported by \cite{Evans2015}: an acoustic mode around 15 kHz becomes unstable preventing the detector functioning for an intracavity power of 50kW, which is much lower than the design power (800 kW). On the other hand, PI hasn’t been observed on AdVirgo detector yet. They are foreseen to appear in the next detection run O4, in two years, when the laser power will be increased and the signal recycling cavity will be installed. Therefore, in order to operate the advanced detectors at their design sensitivity and to realize the next generation of GW detectors, it is necessary to mitigate the PI effects by damping the unstable mirror modes.

\subsection{PI Mitigation Strategies}

A wide variety of techniques have been suggested to overcome PI in GW detectors; one can divide them in two categories: passive and active techniques. The passive techniques aim at preventing the instabilities by changing the cavity parameters of the detector so that the PI no longer occurs. One kind of these techniques acts by increasing the loss of the mirror mechanical modes with passive dampers. Ring dampers \cite{Gras2008} and piezoelectric acoustic mode dampers \cite{Gras2015} have been proposed. These methods are able to decrease the parametric gain, but their effect depends on the frequency of the mechanical mode and care has to be taken on their design to limit the cost in terms of thermal noise. Among them the acoustic mode dampers seem promising \cite{Biscans2018} and have been applied on LIGO mirrors. Another passive technique is the thermal tuning. It consists in changing the radius of curvature of one or more optics by means of thermal actuators \cite{Degallaix2007}. This method has been used during the first observation run of LIGO \cite{Evans2015}. By tuning the resonant frequency of the high order mode involved in the parametric interaction, the amplification of the mechanical mode has been stopped, the intracavity power could be increased up to 100 kW allowing the first detection of GWs. This technique, though, is not enough to avoid all the unstable modes at the nominal power $\approx$ 1MW; because of the high density of the mechanical unstable modes, a detuning which is effective for one mode will bring other modes into resonance.
The active techniques consist in monitoring the onset of the instability and suppress it by a feedback that damps the mirror unstable mode. Miller et al. \cite{Miller2011} proposed the active damping by means of the electrostatic actuators of LIGO mirrors. Recently Blair et al. \cite{Blair2017} proved the effectiveness of this technique. The unstable mirror vibration mode at 15kHz was successfully damped by using the electrostatic control system, normally used to provide longitudinal actuation on the mirrors for the interferometer lock. But this technique is limited by the fixed position of the four actuators. Not all the mechanical modes can be efficiently damped.\\ 
Within the Virgo collaboration different groups are doing research on PI: Universit\`a di Roma Sapienza is involved in simulations and development of passive dampers; Institut Foton in Rennes is studying the issue of PI early detection. A group in LKB is also involved in the PI simulations. Our team in Artemis laboratory is working on the active mitigation solutions. The main goal of our research program is to build a flexible attenuation system for the PI, based on the radiation pressure applied by an auxiliary laser.  

\section{PI Damping System Based on Radiation Pressure}

\subsection{Concept}
\begin{figure}
\centerline{\includegraphics[width=0.8\linewidth]{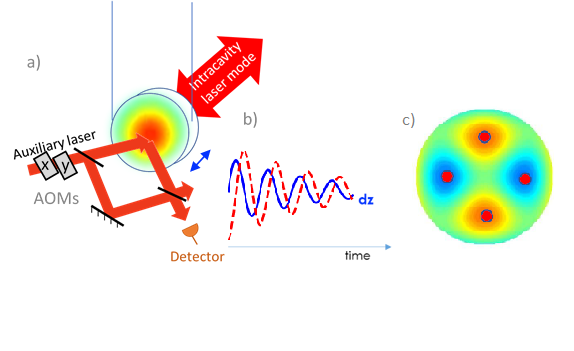}}
\caption[]{PI damping system principle. a) An auxiliary laser is reflected on the suspended end-cavity mirror exerting a radiation pressure force where the mechanical mode induces a deformation of the mirror surface, dz, along the cavity axis. The Mach-Zehnder interferometer allows local detection of the mirror displacements. b) The laser amplitude (red dashed line) is modulated at the same frequency, in phase quadrature to damp the mechanical mode (blue line). This plot is for illustrative purposes only. c) Spatial profile of a higher order mechanical mode, the red dots represent the areas where the radiation pressure of the auxiliary beam must be applied.}
\label{fig:princi}
\end{figure}

Such attenuator device consists of two cascaded acousto-optic modulators (AOM) used as deflectors (for a 2D scan in X and Y direction). An auxiliary laser beam is placed by the AOMs in phase quadrature on the mirror’s areas where the surface deformation, due to the mechanical mode, is maximal. The auxiliary beam amplitude is modulated at the mechanical mode frequency f$_m$ by the AOM itself or by an intensity modulator, it is injected through the mirror’s back surface and reflected on the high reflectivity surface, with a 90$^o$ phase shift with respect to the mechanical mode oscillations. In such a way, it applies a viscous damping force by radiation pressure, meaning that dissipation increases for that particular mode which eventually is damped. Moreover, the auxiliary laser can be used for detection of the unstable modes. For this, the auxiliary beam is split before the contact with the mirror substrate and it is recombined after the reflection on the high reflectivity coating. The output of the Mach-Zehnder interferometer will allow to locally monitor the mirror deformation amplitude, phase and frequency.\\ In the Figure \ref{fig:princi}, a simplified scheme of the PI-damping device is represented (a). Here the mechanical mode induces a longitudinal displacement dz at mirror center (blue straight line in (b)) which is damped by the force applied by the auxiliary laser (red dashed line). If the mechanical mode shows a more complicated spatial profile like in (c), the laser is continuously moved from one lobe to the other.\\
Compared to the active electrostatic damping system, this attenuation device can be more effective since the overlap between the applied force and the spatial shape of the unstable modes can be finely adjusted by the X-Y deflectors. It is also more flexible: the optical power, frequency and phase used for the actuation can be controlled by the AOMs themselves or by an additional intensity modulator. 

\subsection{First Results}
\begin{figure}
\begin{minipage}{0.75\linewidth}
\centerline{\includegraphics[width=\linewidth]{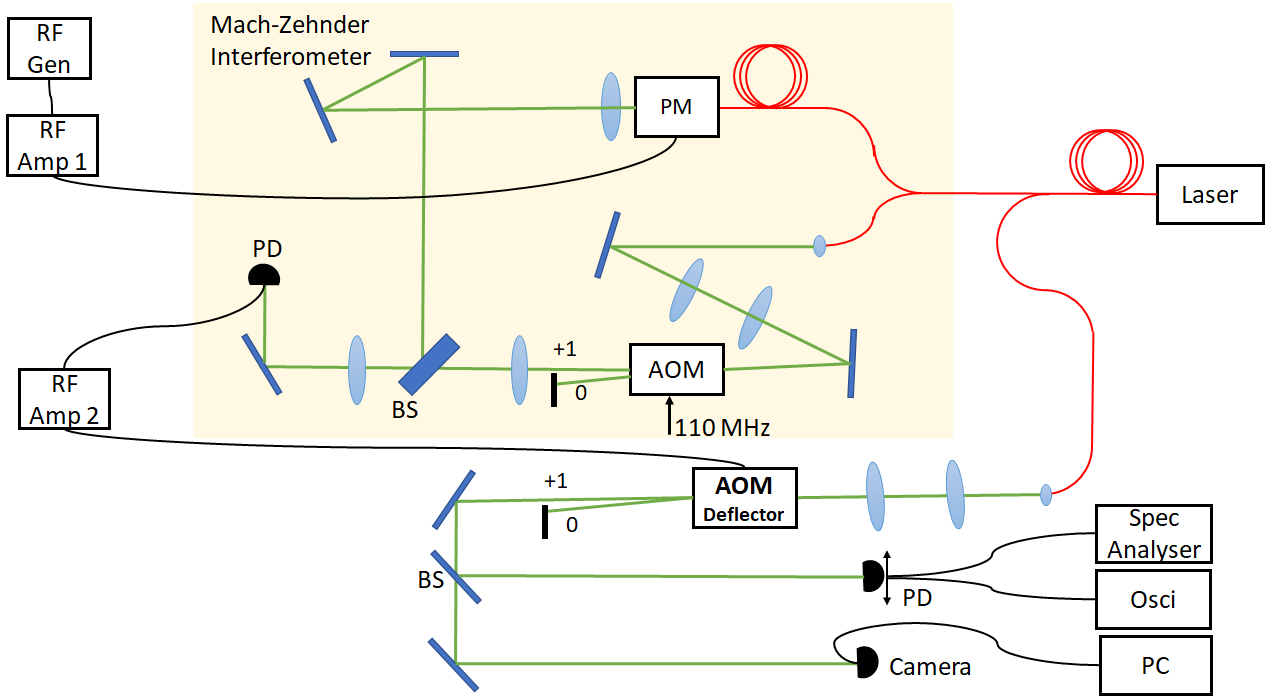}}
\end{minipage}
\begin{minipage}{0.2\linewidth}
\centerline{\includegraphics[width=\linewidth]{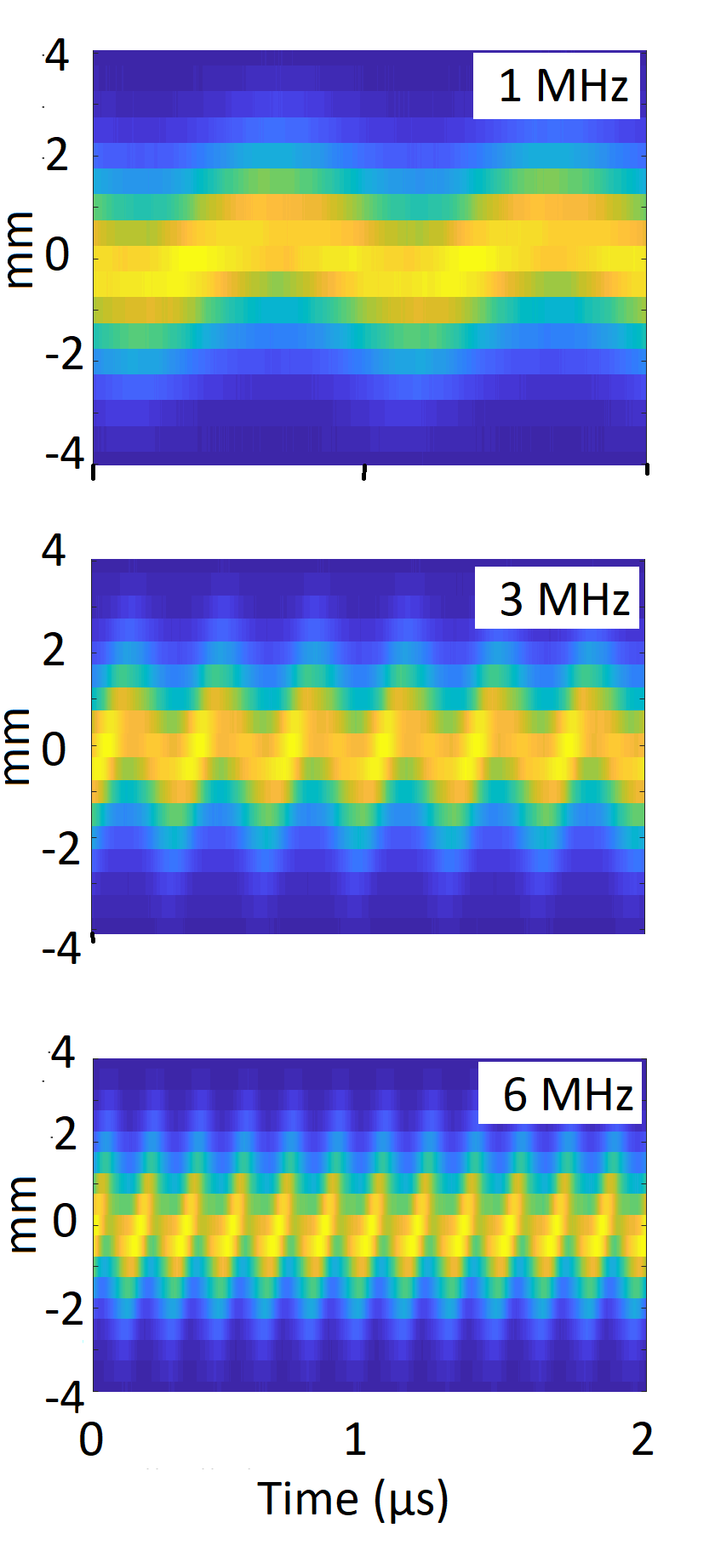}}
\end{minipage}
\caption[]{Left: schematic drawing of the experimental setup. The yellow area is the Mach-Zhender interferometer used for the generation of the RF signal which drives the AOM deflector MT110-A1.5-1064. A Photodetector (PD) on a translation stage is used to monitor the beam displacement. BS: beam Splitter; PM: Phase Modulator. RF Gen: Radio Frequency Generator at frequency f$_m$; RF Amp: Radio Frequency Amplifier; +1: AOM diffraction order.\\Right: Beam displacement as function of time for three values of f$_m$: 1, 3, 6 MHz.}
\label{fig:setup}
\end{figure}

Since November 2018, our team has started the experimental investigation of AOM deflectors in order to define the best characteristics needed for the PI damping purpose. The auxiliary laser should be moved fast in order to hit the mirror on different areas during one period of the mirror mechanical mode. This means that the RF driving frequency of the AOM must change value rapidly. We estimated a scan frequency at 10 MHz is enough for damping 1kHz-100kHz range modes. Moreover the full scan angle should be wide so that a 35 cm wide mirror like the GW detector's ones can be covered.\\  
We have begun by studying the performances of an AOM already available in our laboratory: the model MT110-A1.5-1064 from AA Opto Electronic. Its central driving frequency is 110 MHz, this determines the central position around which the diffracted beam will be moved. It is not trivial to generate radio frequency (RF) signals (for this AOM at 110 MHz) whose frequency is modulated at a rate of 10 MHz or more. In order to do this we exploited the beating of two optical signals as explained below.\\The experimental setup used for the AOM characterization is depicted in the left panel of Figure \ref{fig:setup}.The laser source is a fiber laser with a wavelength of 1064nm amplified to 100mW. The laser is split in three beams, one that goes to the AOM under test, the beam spot size is 300 $\mu$m at the AOM input, this is the smaller accepted beam-size according to the AOM datasheet. The smaller the spot size, the larger is the AOM response bandwidth. The other two beams are used for a Mach-Zhender interferometer. In one arm the laser beam is frequency shifted by 110 MHz by an AOM and in the other arm the phase of the beam is modulated at the frequency f$_m$, which is set from 1 to 10 MHz. The amplitude of the phase modulation $\phi_m$ is set by adjusting the amplitude of the RF generator. The two arms overlap on a Photodiode which detects the beating signal whose frequency varies sinusoidally from 110 MHz - f$_m$ to 110 MHz + f$_m$. 
The photodiode signal is used to drive the tested AOM. We detect the deflected beam with a 1mm wide photodiode on a translation stage, recording the signal every 500 $\mu$m on the transverse plane we are able to track the beam displacement (see right panel of Figure \ref{fig:setup}).
The total scan angle is of 15 mrad, beyond this range, the diffraction efficiency of our AOM drops significantly (not shown). This kind of AOM is thus not suitable for our application. 

\subsection{Perspectives}
The reported results have been used to define the best AOM specifications for the PI attenuation device. Our choice is a trade-off between a big scan-angle and rapidity of the AOM response.The chosen AOM, model 4225-2 from Gooch\&Housego has a scan-angle of 35 mrad for a rather flat efficiency curve and a response time of 33ns. Since the optical power needed for PI attenuation is unknown, the chosen AOM has a high damage threshold allowing up to few tens of watts of injected power. The other key element of the PI damping system based on radiation pressure is the RF generator for driving the AOMs. The Mach-Zhender interferometer shown in this report will be replaced by a software defined radio reconfigurable device from National Instruments which is able to generate a signal in the range 1MHz-6GHz and change frequency within 5 ns. The AOMs will be controlled by a computer program interfaced with the RF device.\\
In the short-term, the auxiliary laser position control will be assembled, including a power stabilization loop in order to overcome the angle dependency of the AOMs efficiency. 

\section{Conclusions}
PI is a harmful phenomenon for GW detectors. We have reported on the research program of Artemis laboratory on PI mitigation. The goal is to develop a flexible device for PI mitigation based on radiation pressure. The device is under development, it will be characterized and then tested. Exploitable results are waited before starting of O4. 

\section*{Acknowledgments}
The research program on PI mitigation based on radiation pressure has received the financial support of Labex First-TF, F\'ed\'eration Doblin, Observatoire de la C\^ote d’Azur, Univ\'ersit\'e de Nice-Sophia Antipolis. We acknowledge European Gravitational Observatory for a three year PhD grant on this topic.

\end{document}